\documentclass[]{article}

\def\sqr#1#2{{\vcenter{\hrule height.#2pt\hbox{\vrule width.#2pt height#1pt
\kern#1pt
\vrule width.#2pt}\hrule height.#2pt}}}

\begin{document}

\title{Potential--density
pairs for axisymmetric galaxies: the influence of scalar fields}

\author{M. A.~ Rodr\'\i guez--Meza$^1$, Jorge L. ~Cervantes--Cota$^1$, \\
  M.I. Pedraza$^2$, J.F. Tlapanco$^2$,
and E.M. De la Calleja$^2$ \\
$^1$Depto. de F\'{\i}sica, Instituto Nacional de Investigaciones
Nucleares, \\ Apdo. Postal 18-1027, 11801 M\'{e}xico D.F. \\
$^2$Instituto de F\'{\i}sica,  Benem\'{e}rita
Universidad Aut\'{o}noma de Puebla, \\
Apdo. Postal J-48, Puebla 72570 M\'{e}xico}
\maketitle
\setcounter{page}{1}

\begin{abstract}
We present a formulation for potential-density pairs to describe axi\-symmetric
galaxies in the Newtonian limit of scalar-tensor
theories of gravity.   The scalar field is described by a modified Helmholtz
equation with a source that is coupled to the standard Poisson equation of
Newtonian gravity.  The net gravitational force is given by two
contributions: the standard Newtonian potential plus a term stemming from
massive scalar fields.  General solutions have been found for axisymmetric
systems and the multipole expansion of the Yukawa potential is given. 
In parti\-cu\-lar,
we have computed potential--density pairs of
galactic disks for an exponential profile and their rotation curves.
\end{abstract}


{Pacs: 04.50.+h, 04.25.Nx, 98.10.+z, 98.62.Gq, 98.62.Hr}

%%%%%%%%%%%%%%%%SECTION%%%%%%%%%%%%%%%%%%%%%%%%
\section{Introduction}
In recent years there has been much interest in the study of
dark matter and energy in a cosmological and astrophysical context.  This has
been motived by recent independent observational data in the Cosmic Microwave
Background Radiation (CMBR) at various angular
scales \cite{deB00},  type Ia supernovae \cite{Ri98}, and
the 2dF Galaxy Redshift survey \cite{Pe02},  among others. These suggest that
$\Omega = \Omega_{\rm \Lambda} + \Omega_{\rm m} \approx 1$, with
$\Omega_{\rm \Lambda} \approx 0.7$ and $\Omega_{\rm m} \approx 0.3$,
implying the existence of dark energy and dark matter, respectively. In
this way galaxies are expected to possess dark components and, in 
accordance with the
rotation curves of stars and gas around the centers of spirals, they
might be in the form of halos \cite{OsPe73} and  must contribute to at
least 3 to 10 times the mass of the visible matter \cite{KoTu90}. The 
origin of these
dark components is, however, unknown.

The theoretical framework to explain the existence of dark components
finds its origin in theories of elementary particles physics, with the addition
of the action of gravity.  There exist many theories (grand unification
schemes, string theories, braneworlds, etc) that involve such physics, but
scalar--tensor theories (STT) of gravity are typically found to represent
classical effective descriptions of such original theories \cite{Gr88}.  In
this way, the scalar fields of these theories are the natural 
candidates to be the
quintessence field \cite{Ca98}, as a remnant of some cosmological
function.  It has  been even suggested that the quintessence field is 
the scalar field that
also acts on local planetary scales \cite{Fu00} or on galactic
scales \cite{MaGu01}.  Moreover, massive scalar fields might account to the
dark matter components of galaxies in the form of halos.

Motivated by the above arguments, we have recently studied some STT
effects in galactic systems\cite{RoCe2004}.  We have computed
spherical potential--density pairs \cite{BiTr94} coming from such 
theories in their Newtonian
approximation. In the present report we compute general 
potential--density pairs
and take some limiting cases. Then, we consider the general 
axisymmetric case and
take, as an example, an exponential profile, that typically describes 
stars on the
disk. Also, rotation curves of such a model can be directly obtained.

This paper is organized as follows: in section 2 we present the 
equations of the
Newtonian approximation of a general scalar--tensor theory of gravity and
solutions are given in terms of integrals of Green
functions. In section 3, we discuss solutions for point-like mass 
distributions and
show how a multipole expansion can be done for the Newtonian limit
contribution of the scalar field.
In section 4, computations are done for general axisymmetric
potentials and rotation velocities of stars in galaxies. Then,  an 
exponential disk
is considered.  In section 5, we present our conclusions.

%%%%%%%%%%%%%%%%SECTION%%%%%%%%%%%%%%%%%%%%%%%%
\section{Scalar Fields and the Newtonian Approximation}
In references \cite{RoCe2004,He91} are derived the following 
differential equations,
that are valid for a general scalar--tensor theory in its Newtonian limit,
\begin{eqnarray}
\nabla^2 \psi &=& 4\pi \rho \; , \label{pares_eq_psi} \\
\nabla^2 \bar{\phi} - m^2 \bar{\phi} &=& - 8\pi \alpha\rho \; ,
\label{pares_eq_phibar}
\end{eqnarray}
in which we consider deviations of the scalar field $\bar{\phi}$ from 
some average
value, given by the inverse of the Newtonian constant, $1/G$; 
$\alpha$ is a some
free constant of the theory and $m$ the mass of some boson particle, see
details in \cite{RoCe2004}.

The standard Newtonian potential, $\psi$, is obtained when the 
perturbation is set
to zero, $\bar{\phi}=0$ and $\alpha=0$. Otherwise, the new Newtonian
potential is given by
\begin{equation}
\Phi_N  = \psi - \frac{1}{2} \bar{\phi} \, .
\end{equation}

The next step is to find solutions for this new Newtonian potential given
a density profile, that is, to find the so--called potential--density 
pairs\cite{BiTr94}.
General solutions to Eqs. (\ref{pares_eq_psi}) and (\ref{pares_eq_phibar})
can be found in terms of the corresponding Green functions
\begin{eqnarray}
\psi &=& -\int d{\bf r}_s \frac{\rho({\bf r}_s)}{|{\bf r}-{\bf r}_s|}
+ \mbox{B.C.} \label{pares_eq_gralpsi} \; , \\
\bar{\phi} &=& 2\alpha\int d{\bf r}_s
\frac{\rho({\bf r}_s)  {\rm e}^{-m |{\bf r}-{\bf r}_s|}}
	{| {\bf r}-{\bf r}_s|} + \mbox{B.C.} \label{pares_eq_gralphi} \; ,
\end{eqnarray}
and the new Newtonian potential is
\begin{eqnarray}
\Phi_N &=&  \psi
- \frac{1}{2} \bar{\phi} = - \int d{\bf r}_s
\frac{\rho({\bf r}_s)}{|{\bf r}-{\bf r}_s|}
%\nonumber \\ &&
-\alpha\int d{\bf r}_s \frac{\rho({\bf r}_s)
{\rm e}^{- m |{\bf r}-{\bf r}_s|}}
{| {\bf r}-{\bf r}_s|} + \mbox{B.C.} \label{pares_eq_gralPsiN}
\end{eqnarray}
The first term of Eq. (\ref{pares_eq_gralPsiN}), given by $\psi$, is the
contribution of the usual Newtonian gravitation (without scalar
fields), while information about the scalar field is contained in the
second term, that is, arising from the influence function determined by the
Helmholtz Green function, where the coupling $\alpha$ enters
as part of a source factor.

%%%%%%%%%%%%%%%%SECTION%%%%%%%%%%%%%%%%%%%%%%%%
\section{Point--like masses and multipole expansion}
Now we present solutions for point--like masses and the multipole expansion of
the potentials which are useful for numerical simulations of galaxies.
Substituting the following
distribution of $N$ point-like masses in Eq. (\ref{pares_eq_gralPsiN})
\begin{equation}
\rho({\bf r}) = \sum_{s=1}^N m_s \delta({\bf r}-{\bf r}_s)
\end{equation}
one obtains \cite{Fu71,FiTa99}
\begin{eqnarray}
\Phi_N &=&  -\sum_{s=1}^N
\frac{m_s}{|{\bf r}-{\bf r}_s|}
%\nonumber \\
%&&
-\alpha\sum_{s=1}^N \frac{m_s
{\rm e}^{- |{\bf r}-{\bf r}_s|/\lambda}}
{| {\bf r}-{\bf r}_s|} \label{pares_eq_pointlike}
\end{eqnarray}
with $m_s$ a source mass at spatial position ${\bf r}_s$, and the 
total gravitational
force on a particle of mass $m_i$ is
\begin{equation}
\sum {\bf F} = - \nabla \Phi_N = m_i {\bf a} \, ,
\end{equation}
where $\lambda = \hbar/mc$ is the Compton wavelength of the effective 
mass ($m$) of
some elementary particle (boson) determined by specific particle 
physics models. In
what follows we will use $\lambda$ instead of $m^{-1}$.
This length can have a range of values depending on particular
particle physics models.  The first right hand term in 
Eq.(\ref{pares_eq_pointlike}) is the pure
Newtonian part and the second one is the dark matter contribution 
which is of the Yukawa type.
There are two limits: On the one hand, for if
$r\gg \lambda$ (or $\lambda \rightarrow 0$)
one recovers the Newtonian theory of gravity. On the other hand, for if
$r \ll \lambda$ (or $\lambda \rightarrow \infty$) one again obtains the
Newtonian theory, but now with a rescaled Newtonian
constant, $G \rightarrow G_{N} (1+\alpha)$. There are
stringent constraints on the possible $\lambda$--$\alpha$ values determined
by measurements on  local scales \cite{FiTa99}.

In the past the above solutions have been used to solve
the missing mass problem in spirals \cite{Sa84,Ec93} as an alternative to
considering a distribution of dark matter.  This was done assuming that
most of the galactic mass is located in the galactic center, and then
considering the center to be a point source.  In our present investigation we
do not avoid dark matter, since our model predicts that bosonic dark matter
produces, through a scalar field associated to it, a modification of
Newtonian gravity theory.  This dark matter is presumably clumped in the
form of dark halos.   Therefore we will consider in what follows that
a dark halo is axisymmetrically distributed along an observable spiral and
beyond, having some density profile.  Next, we compute the
potentials, and some astrophysical quantities, for general halo density
distributions.

The standard gravitational potential due to a distribution of mass 
$\rho({\bf r})$,
in a point exterior to the distribution and without considering the boundary
condition, can be expanded as \cite{Jackson}
\begin{equation}
\psi({\bf r}) = - \sum_{l=0}^\infty \sum_{n=-l}^l \frac{4\pi}{2l+1} q_{ln}
\frac{Y_{ln}(\theta,\varphi)}{r^{l+1}} \, ,
\end{equation}
with the expansion
\begin{equation}
\frac{1}{|{\bf r}-{\bf r}_s|}=4\pi \sum_{l=0}^\infty \sum_{n=-l}^l 
\frac{1}{2l+1}
\frac{r_<^l}{r_>^{l+1}} Y_{ln}^*(\theta',\varphi') Y_{ln}(\theta,\varphi),
\end{equation}
where $r_<$ is the smaller of $|{\bf r}|$ and $|{\bf r}_s|$, and
$r_>$ is the larger of $|{\bf r}|$ and $|{\bf r}_s|$; 
$Y_{ln}(\theta,\varphi)$ are the spherical harmonics.

Here we are interested in the potential outside the distribution of 
mass, consequently, the
coefficients of the expansion of $\psi$, and known as multipoles are given by
\begin{equation}
q_{ln}= \int d{\bf r}' Y_{ln}(\theta',\varphi') r'^l \rho({\bf r}')  \, .
\end{equation}

In the case of the scalar field, with the expansion
\begin{equation}
\frac{\exp(-m |{\bf r}-{\bf r}_s|)}{|{\bf r}-{\bf r}_s|}=
4\pi m \sum_{l=0}^\infty \sum_{n=-l}^l 
i_l(m r_<) k_l(m r_>) Y_{ln}^*(\theta',\varphi') Y_{ln}(\theta,\varphi),
\end{equation}
the contribution of the scalar field to the Newtonian gravitational potential
can be written as
\begin{equation}
\frac{1}{2\alpha} \bar{\phi}({\bf r}) =
\sum_{l=0}^\infty \sum_{n=-l}^l \frac{4\pi}{2l+1} \bar{q}_{ln}  k_l(m r)
Y_{ln}(\theta,\varphi),
\end{equation}
where $i_l(x)$ and $k_l(x)$ are the modified spherical Bessel functions and we
have defined the multipoles for the scalar field as
\begin{equation}
\bar{q}_{ln} = m \int d{\bf r}' \, Y_{ln}(\theta',\varphi')\, i_l(m 
r')\, \rho({\bf r}') \, .
\end{equation}

%%%%%%%%%%%%%%%%SECTION%%%%%%%%%%%%%%%%%%%%%%%%
\section{General axisymmetric density distributions and their 
rotation velocities}
Given that we are interested in axisymmetric galaxies in what follows
we will only consider the case of this symmetry. Additionally, we use
flatness boundary conditions (B.C.) at infinity, such that the boundary terms
in Eqs. (\ref{pares_eq_gralpsi})-(\ref{pares_eq_gralphi}) are zero.  Moreover,
regularity conditions must be applied to spatial points where the potentials
are singular. For axisymmetric systems these conditions mean that
$d \psi/dr=d \bar{\phi}/dr=0$ along the symmetry axis.  Accordingly,
we assume that
\begin{equation}
\rho(r,z) = \left\{
\begin{array}{lll}
\rho(r,z) & ; & l < L \\
0 & ; & l \ge L
\end{array}
\right.
\end{equation}
where we define the boundary surface by a single-valued function $r=Z(z)$
with $z$ in the interval $(z_0,z_1)$
and $l=\sqrt{r^2+z^2}$, and the same relation for $L$ with $(r,z)$ on the
boundary surface.
We obtain for $\psi$
\begin{equation}\label{pares_eq_finalpsi}
\psi(r,z) = - \int_{z_0}^{z_1} dz \int_0^{Z(z)} r_s dr_s\; \rho(r_s,z_s)
\int_0^\infty dk\; I_0(kr_<) K_0(kr_>)  \cos k(z-z_s)
\; ,
\end{equation}
and for $\bar{\phi}$
\begin{equation}\label{pares_eq_finalphi}
\frac{1}{2}\bar{\phi}(r,z) = -\alpha \int_{z_0}^{z_1} dz 
\int_0^{Z(z)} r_s dr_s\; \rho(r_s,z_s)
\int_0^\infty dk\; I_0(\nu r_<) K_0(\nu r_>)  \cos k(z-z_s)
\; ,
\end{equation}
where $I_0(x)$ and $K_0(x)$ are modified Bessel functions of index $n=0$
and $\nu^2 = k^2 + \lambda^{-2}$.
Then, the
new Newtonian potential can be  obtained using (\ref{pares_eq_gralPsiN}).

An important quantity that characterizes the steady state of astrophysical
systems is the circular velocity.
The circular velocity for a test particle moving on the equatorial plane
of an axisymmetric galactic system is given by
\begin{eqnarray} \label{pares_eq_vc2_1}
v_c^2 &=& r \frac{d\Phi_N}{dr} \; .
\end{eqnarray}

We may consider as an example that the distribution of mass is
\begin{equation}
\rho(r,z)=A \exp(-r/r_0) \delta(z) \; ,
\end{equation}
where $r_0$ is the effective radial extension of the system. This 
distribution of mass
is typically observed in stars of a disk galaxy. In this case the 
Newtonian potential is
\begin{eqnarray}
\hspace{-0.15in}
\Phi_N(r) &=& - A  \left\{
\int_0^{r} r_s dr_s\; \exp(-r_s/r_0)
\int_0^\infty dk\; I_0(kr_s) K_0(kr)  \cos k(z)
\right. \nonumber \\ &&
+\int_r^{\infty} r_s dr_s\; \exp(-r_s/r_0)
\int_0^\infty dk\; I_0(kr) K_0(kr_s)  \cos k(z)
\nonumber \\ &&
+\alpha \int_0^{r} r_s dr_s\; \exp(-r_s/r_0)
\int_0^\infty dk\; I_0(\nu r_s) K_0(\nu r)  \cos k(z)
\nonumber \\ && \left.
+\alpha \int_r^{\infty} r_s dr_s\; \exp(-r_s/r_0)
\int_0^\infty dk\; I_0(\nu r) K_0(\nu r_s)  \cos k(z)
\right\}  .
\end{eqnarray}
Using Eq. (19) the rotation curve is obtained.

%%%%%%%%%%%%%%%%SECTION%%%%%%%%%%%%%%%%%%%%%%%%
\section{Discussion and Conclusions}
We have found potential--density pairs of axisymmetric galactic systems within
the context of linearized scalar--tensor theories of gravitation.  The
influence of massive scalar fields is given by $\bar{\phi}$, determined by
Eq. (\ref{pares_eq_finalphi}). Circular velocities of {\it stars} can 
be directly obtained
in the axisymmetric system; by stars
we mean probe particles that follow the dark halo potential.
Specifically, these results were used to find potential--density pairs
for an exponential profile.    In general the contribution due to 
massive scalar
fields is non--trivial, see for instance Eq. (21), and interestingly, forces
on circular orbits of stars depend on the parameters $\lambda$ and
$\alpha$ in a rather complicated way.  This means that even when local
experiments force $\alpha$ to be a very small number \cite{FiTa99}, the
amplitude of forces exerted on stars is not necessarily very small and may
contribute significantly to the dynamics of stars.  Alternatively, 
one may interpret
the local Newtonian constant as given by $ (1 + \alpha) \langle \phi 
\rangle^{-1} $,
instead of being given by $\langle \phi \rangle^{-1}$ ($1$ in our 
convention). In this case,
the local measurement constraints are automatically satisfied, and at
scales larger than $\lambda$ one sees a reduction of $1/(1+\alpha)$ in the
Newtonian constant.

In the past, different authors have used  point solutions, Eq. (6), 
to solve the missing
mass problem encountered in the rotation velocities of spirals
and in galaxy cluster dynamics \cite{Sa84,Ec93}. These models were used
as an alternative to avoid dark matter.
Indeed, for single galaxies one can adjust the
parameters ($\alpha$, $\lambda$) to solve these problems without
the need for dark matter.  However, these models do not provide a good
description of the systematics of galaxy rotation curves because they
predict the scale $\lambda$ to be independent of the galactic 
luminosity, and this conflicts
with observations for different galactic sizes \cite{Ag01} unless one assumes
various $\lambda$'s, and hence various fundamental masses, $m$, one for each
galaxy size. Such a particle spectrum is not expected from theoretical
arguments; it represents a considerable fine tuning of masses.  This criticism
would also apply to our models.  The purpose of the
present investigation was, however, not to present a model alternative
to dark matter in order to solve the missing mass problem, but to 
compute the influence
of scalar--field dark matter distributed in the form of an 
axisymmetric dark halo.  This
contribution, together with the other components of the galaxy, will 
give rise to a flat
velocity curve, in a similar way as in Newtonian mechanics.

\bigskip

{\Large\bf Acknowledgments}
\bigskip

This contribution was written in honor of the 60th birthday of Prof. 
Alberto Garc{\'i}a, who have taught us the importance of finding out 
a physical meaning to the exact solutions in gravitation.

This work was supported in part by CONACYT, grant number 44917-F.

%%%%%%%%%%%%%%%%%%%%%%%%%%%%%%%%%%%%%%%%%%%%%%%%%
%                      Bibliography                        %
%%%%%%%%%%%%%%%%%%%%%%%%%%%%%%%%%%%%%%%%%%%%%%%%%

\end{document}